\documentclass[runningheads]{llncs}

\usepackage{graphicx}
\usepackage{url}
\usepackage{cite}
\usepackage{bbding}
\usepackage{enumitem}

\begin{document}
\title{MIDI-Sandwich: Multi-model Multi-task Hierarchical Conditional VAE-GAN networks for Symbolic Single-track Music Generation\thanks{Supported by the National Key Research and Development Program of China under Grant 2016YFB1000403.}}

%
%
\author{Xia Liang\inst{1 (}\Envelope\inst{)} \and
Junmin Wu\inst{1}\and
Yan Yin\inst{1}}
\authorrunning{F. Author et al.}
%
\institute{University of Science and Technology of China, Hefei, Anhui, China
\email{sa517190@mail.ustc.edu.cn}\\
}

\maketitle              

\begin{abstract} 
Most existing neural network models for music generation explore how to generate music bars, then directly splice the music bars into a song. However, these methods do not explore the relationship between the bars, and the connected song as a whole has no musical form structure and sense of musical direction. To address this issue, we propose a Multi-model Multi-task Hierarchical Conditional VAE-GAN (Variational Autoencoder-Generative adversarial networks) networks, named MIDI-Sandwich, which combines musical knowledge, such as musical form, tonic, and melodic motion. The MIDI-Sandwich has two submodels: Hierarchical Conditional Variational Autoencoder (HCVAE) and Hierarchical Conditional Generative Adversarial Network (HCGAN). The HCVAE uses hierarchical structure. The underlying layer of HCVAE uses Local Conditional Variational Autoencoder (L-CVAE) to generate a music bar which is pre-specified by the First and Last Notes (FLN). The upper layer of HCVAE uses Global Variational Autoencoder(G-VAE) to analyze the latent vector sequence generated by the L-CVAE encoder, to explore the musical relationship between the bars, and to produce the song pieced together by multiple music bars generated by the L-CVAE decoder, which makes the song both have musical structure and sense of direction. At the same time, the HCVAE shares a part of itself with the HCGAN to further improve the performance of the generated music. The MIDI-Sandwich is validated on the Nottingham dataset and is able to generate a single-track melody sequence (17x8 beats), which is superior to the length of most of the generated models (8 to 32 beats). Meanwhile, by referring to the experimental methods of many classical kinds of literature, the quality evaluation of the generated music is performed. The above experiments prove the validity of the model.


\end{abstract}  
\section{Introduction}
Algorithmic composition is not a new idea. Earlier, there was the use of statistical methods to create music\cite{conklin2003music}. Research on automatic music generation has made great progress due to the development of deep neural networks\cite{dai2018music}\cite{fessahaye2019t}\cite{nam2019deep}.

Compared with computer vision (CV), natural language processing (NLP) and other common fields of deep learning, automatic music generation has some inherent challenges\cite{briot2017deep}, such as the complexity of music sequence itself in structure, the difficulty of objective evaluation of music generation results. Our approach is to use the symbol (MIDI format) music format for automatic music generation. The model based on Recurrent Neural Network (RNN) is the most common for symbolic-domain music generation, such as Boulanger et al.\cite{boulanger2012modeling}'s RNN-based model that can learn harmonic and rhythmic probabilistic rules, Google Magenta\cite{casella2001magenta}. As well as, Waite et al.\cite{waite2016project} use Long Short-Term Memory (LSTM) to make music. The RNN-based method is relatively mature but has some limitations: The music output of the model is limited in length, and a song which is directly spliced by some fragments is not good as a whole; Most models are difficult to continue to develop.

Xiaoice Band\cite{zhu2018xiaoice} is a model that deals with multi-track music. It introduces the concept of music, such as chord progression and rhythm shape. However, a generated music do not reflect the relationship between the bars, and data relies on multiple tracks and chords. Midi-Net also adds chords to the model\cite{yang2017midinet}. It proposes a novel conditional mechanism to exploit available prior knowledge. Pi from song\cite{chu2016song} is based on chords to generate each section but lacks control constraints between the individual bars. The above methods of using chords to limit the music sequence have a problem of having a chord music format dependency. For this, this paper adopts the First and Last Notes (FLN) to replace the chord progression as a constraint. Therefore in the format of only the main melody track, it is possible to train and generate music with a melodic motion. As a result, it reduces dependency on the data format and improves the applicability of the model.

Using RNN to construct Generative adversarial networks (GAN) will not be able to use traditional gradient descent to optimize. SeqGAN\cite{yu2017seqgan} solves this problem with Monte Carlo tree search of Reinforcement Learning. The model works to some extent, but the Monte Carlo search greatly increases the training time as the sequence length increases. MuseGAN\cite{dong2018musegan} is a GAN model by using CNN to construct which can generate 4-bar multi-track music phrases. C-RNN-GAN\cite{mogren2016c} proposes a generative adversarial model with both Generator (G) and Discriminator/Critor (D) being LSTMs that works on continuous sequential data. MusicVAE\cite{roberts2017hierarchical} is a generative recurrent Variational Autoencoder (VAE) model that can generates melodies and drum beats.Transformer\cite{huang2018music} applies the latest Attention technology to produce music which has a longer sequence. MusicVAE and Transformer do complete the extension of the sequence length, but the model is not very interpretable, and there is no fusion musical knowledge. Yang et al.\cite{yang2019inspecting} propose a new method to inspect the pitch and rhythm interpretations of the latent representations with the VAE model. At the same time, there is also a practice of merging GAN with VAE that uses the decoder of the VAE as the generator of GAN\cite{akbari2018semi}.

Due to the problems mentioned above, we find that most of the models have high requirements on data format and just generate music fragments instead of a song. In this regard, we propose the MIDI-Sandwich for generating long-term single track symbolic music sequences that have the characteristics of musical form structure and melodic motion. Training can also be done on the dataset with only the single melody track. MIDI-Sandwich is a multi-model multi-task hierarchical architecture, which can be expound from three aspects: 1) Multi-model. We propose four models of HCVAE, HCGAN, Local Contitional VAE (L-CVAE), and Global VAE (G-VAE). Simultaneously, L-CVAE and G-VAE are sub-models of HCVAE, and HCVAE also shares decoder and part of encoder with HCGAN; 2) Hierarchical architecture. The underlying model, the L-CVAE, processes local information, while the upper model, the G-VAE, controls global trends. Compared with MusicVAE and Transformer, our hierarchical model not only increases the length of generated sequences but also practices the control of structure (musical form) and melody direction;  3) Multi-task. The process of training model can be subdivided into three tasks: training L-CVAE, training HCVAE and training HCGAN.


In summary, we make the following contributions:

1. It is first proposed to use the First and Last Notes (FLN) as constraint conditions to restrain the generation of melody. Compared with using chord as a constraint condition in other model, using the FLN as the constraint condition can not only reflect the melodic motion so that the generated melody has a sense of direction, but also reduce the need for data format. It is not necessary to rely on a chord, and the single track can be trained. Since the FLN are related to the tonic and tonality in musical theory, specifying the FLN can maintain a stable tonality of a song. 

2. We use a hierarchical model. The upper layer completes the structural control of music by analyzing the latent vector distribution of music fragments generated by the underlying layer.

3. We propose a multi-model multi-task hierarchical architecture for symbolic single-track music generation. This architecture makes the training process easier to monitor and control and makes the overall model more flexible to expand.

4. We combine HCVAE and HCGAN to smooth the model training curve to improve the quality of generating music, which not only makes the HCVAE decoder as HCGAN generator but also shares the part of the HCVAE encoder with the HCGAN critor. It is a new attempt to integrate VAE and GAN on the issue of automatic music generation. 
	

\section{Background}
\subsection{VAE,VRAE,CVAE} 
Variational Auto-encoder(VAE)\cite{kingma2013auto} consists of an encoder and a decoder. A disadvantage of VAE is that, because of the injected noise and imperfect elementwise measures such as the squared error, the generated samples are often blurry. The VAE loss fuction is defined as follows: 
\vspace*{-0.4\baselineskip}
\begin{equation}
L=E_{X\sim \bar{p} (X)}[E_{z\sim p(Z|X)}[-\ln q(X|Z)]+KL(p(Z|X)||q(Z))]
\vspace*{-0.2\baselineskip}
\end{equation}
where the first and second terms are the reconstruction loss and a prior regularization term that is the Kullback-Leibler (KL) divergence, respectively.

VRAE\cite{bowman2015generating} uses an RNN-based neural network unit instead of a CNN-based unit to build encoder and decoder.
Conditional Variational Autoencoder (CVAE)\cite{sohn2015learning} is an extension of VAE which can generate some data specificed by user.

\subsection{GAN,WGAN,WGAN-GP} 
Generative Adversarial Network (GAN)\cite{goodfellow2014generative} simultaneously trains two models: a generative model to synthesize samples, and a discriminative model (critor) to differentiate between natural and synthesized samples. The training procedure can be formally modeled as a two-player minimax game between the generator G and the discriminator D:
\vspace*{-0.4\baselineskip}
\begin{equation}
\mathop{min}\limits_{G}\mathop{max}\limits_{D}E_{x\sim pd}[\log(D(X))]+E_{z\sim pz}[1-\log(D(G(z)))]
\vspace*{-0.2\baselineskip}
\end{equation}
where $pd$ and $pz$ represent the distribution of real data and the prior distribution of $z$.

However, GAN training is difficult to free from training instability and low-quality generated samples. Different methods have been proposed to prove GAN. For example, WGAN\cite{arjovsky2017wasserstein}, DCGAN\cite{radford2015unsupervised}, WGAN-GP\cite{gulrajani2017improved} deal with the collapse of the model, so that training tends to stabilize. We experiment with the WGAN-GP model which is proposed for some problems with WGAN. WGAN-GP uses a gradient penalty to improve the problem of extreme parameters in weight clipping. The objective function of the differential equation is defined as follows:
\vspace*{-0.4\baselineskip}
\begin{equation}
E_{x\sim pd}[D(x)]-E_{z\sim pz}[D(G(z))]+E_{\hat{x}\sim p\hat{x}}[(\nabla_{\hat{x}}||\hat{x}||-1)^2]
\vspace*{-0.2\baselineskip}
\end{equation}
where $p\hat{x}$ is defined sampling uniformly along straight lines between pairs of points sampled from $pd$ and $pg$, the model distribution. The WGAN-GP model has faster convergence for better optimal solutions and requires less parameter adjustment. 

\section{MIDI-Sandwich}
\begin{figure}
	\begin{minipage}[t]{0.5\linewidth}
		\centering
		\includegraphics[width=1.5in]{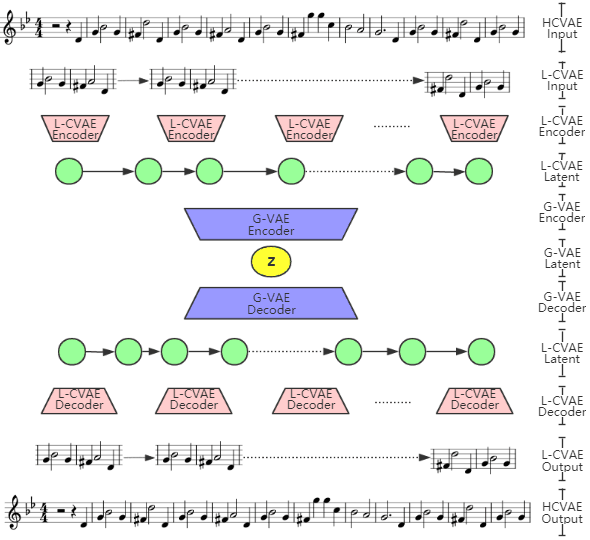}
		\caption{Hirerarchical Conditional Variational Autoencoder (HCVAE)}
		\label{fig:side:a}
	\end{minipage}%
	\begin{minipage}[t]{0.5\linewidth}
		\centering
		\includegraphics[width=1.5in]{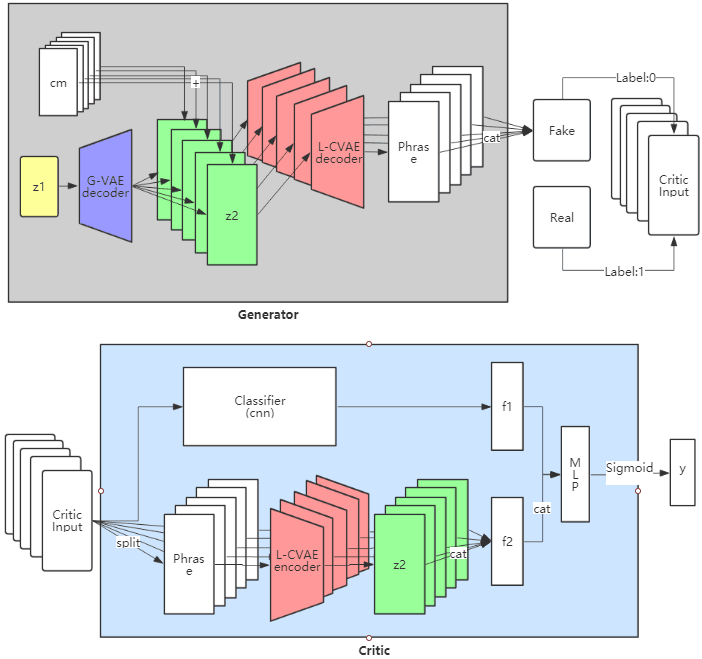}
		\caption{Hirerarchical Conditional Generative Adversarial Network (HCGAN)}
		\label{fig:side:b}
	\end{minipage}
\vspace*{-1.5\baselineskip}
\end{figure}

In this section, we introduce the Multi-model Multi-task Hierarchical Conditional VAE-GAN networks, called MIDI-Sandwich. As shown in Figure \ref{fig:side:a} and Figure \ref{fig:side:b}, the Midi-Sandwich contains two parts: Hierarchical Conditional VAE (HCVAE) and Hierarchical GAN (HCGAN). HCVAE can also be decomposed into two submodels: Local Condition VAE (L-CVAE) and Global VAE (G-VAE). 

Model training can be divided into three tasks: 1) Train Local Condition VAE (L-CVAE); 2) Train HCVAE. G-VAE is a upper layer of HCVAE, while decoder and encoder of L-CVAE are integrated into HCVAE as the bottom layer. 3) Train HCGAN. We use the complete part of HCVAE decoder as the HCGAN generator, and we integrate the part of HCVAE encoder into the Critor of HCGAN. The multi-task mode makes the whole training process more controllable, more explanatory and easier to expand functions than the end-to-end model.

Compaired with CVAE-GAN or other common methods, MIDI-Sandwich has different ways to combine GAN with VAE. It not only makes VAE decoder as GAN generator but also shares parts of VAE encoder with GAN Critor. We do this for two benefits: 1) Smooth the learning curve; 2) Improve the quality of generated results. 
\vspace*{-1.2\baselineskip}

\subsection{Hirerarchical Conditional VAE}

HCVAE consists of hierarchical encoder and hierarchical decoder based on RNN, and hierarchical decoder/encoder is layered as L-CVAE decoder/encoder and G-VAE decoder/encoder. As shown in Figure \ref{fig:side:a}, we first divide a song into several music sections, and L-CVAE decoder is used to map each music section into an latent vector, so as to obtain an latent vector sequence. Then G-VAE encodes, samples, and decodes the latent vector sequence to analyze the manifestation of musical form structure in the latent vector level. Finally, latent vector sequence regenerated by G-VAE decoder is used as the input for L-CVAE decoder to generate music bars which are then composed into a song. We train the L-CVAE first (Task 1), whereafter, we add the trained L-CVAE to HCVAE, then we train the HCVAE (Task 2).

\subsubsection{Local Condition VAE} 
\begin{figure}
	
	\begin{minipage}[t]{0.5\linewidth}
		\centering
		\includegraphics[width=1.4in]{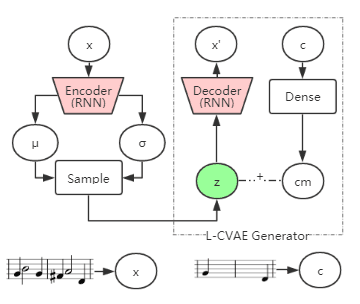}
		\caption{Task1 model: L-CVAE}
		\label{fig:side:c}
	\end{minipage}%
	\begin{minipage}[t]{0.5\linewidth}
		\centering
		\includegraphics[width=1.4in]{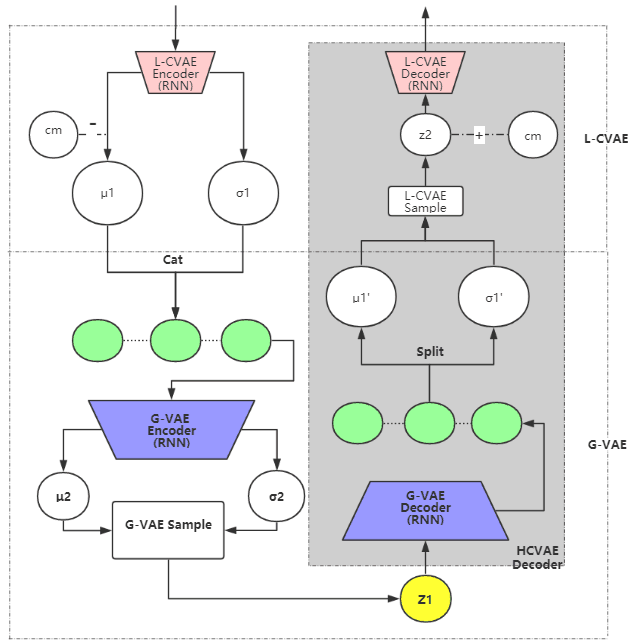}
		\caption{Task2 model: HCVAE}
		\label{fig:side:d}
	\end{minipage}
\vspace*{-1.5\baselineskip}
\end{figure}

On the basis of VAE, we propose the L-CVAE network which uses the First and last notes (FLN) as the constraint conditions for dealing with music fragments. The network structure diagram is shown in Figure \ref{fig:side:c}.

First, we count and sort the frequency of the FLN that appeares in each music fragment in the data set. We take top K higher frequency in the list of FLN and regard them as K classes. This transforms the conditional constraint problem into the K classification constraint problem.

Based on the original VAE model, a dense layer is added to learn the mapping of class labels $c$ to class mean $cm$. This class means $cm$ is added to the latent vector $z$, and the result is used as the input to the decoder. The loss function of L-CVAE is defined as follows:
\vspace*{-0.5\baselineskip}
\begin{equation}
L_{\mu,\sigma^2}=\frac{1}{2}\sum_{i=1}^{d}[(\mu_{(i)}-\mu_{(i)}^c)^2+\sigma_{(i)}^2-\log \sigma_{(i)}^2-1]
\vspace*{-0.5\baselineskip}
\end{equation}
where $\mu_{(i)}^c$ is the learned class mean $cm_{i}$ (cm in Fig \ref{fig:side:c}). The purpose of this is to make the latent vector z of music bar of the same constraint class approximate the mean of this class. Thus, the way to generate the music fragments of the specified constraint is to set the input $z$ of the decoder to the $cm_{i}$, which is the class mean of the specified constraint, plus a Gaussian random errors, then let the L-CVAE decoder generate automatically.
\vspace*{-1.2\baselineskip}

\subsubsection{Global VAE} 
As shown in Figure \ref{fig:side:d}, the Global VAE (G-VAE) is the upper-level model of HCVAE for analyzing the latent vector sequence (The blue circle in Figure \ref{fig:side:d}) encoded by L-CVAE, discovering the implied relation between music section, and rebuilding/generating the latent vector sequence as input subsequently of the L-CVAE decoder.

There are three points to note here:

1. Model training needs to freeze the L-CVAE decoder and encoder. We only train the G-VAE part in Task2;

2. The VAE process is encoding, sampling, and decoding, and the latent vector of the VAE is generated after sampling. The sample of L-CVAE is defined as follows:
\vspace*{-0.5\baselineskip}
\begin{equation}
z = \mu(X)+\sigma^\frac{1}{2}(X)*\epsilon
\vspace*{-0.5\baselineskip}
\end{equation}
where $\epsilon \sim N(0,1)$. The L-CVAE sampling is arranged after G-VAE decoding rather than before G-VAE encoding. The first reason is to avoid random noise. When sampling, $\epsilon$ is introduced  into latent vector as a random error subject to Gaussian distribution, which would interfere with the subsequent latent vector analysis of G-VAE; The second reason is to improve the quality of output. A VAE decoder works best when the input is subject to Gaussian distribution. Reconstructed/Generated latent vectors generated by G-VAE encoder do not necessarily subject to Gaussian distribution exactly, so the input of the L-CVAE decoder is subject to Gaussian distribution forcibly to improve the generation effect when the sampling is arranged after the decoding of G-VAE;

3. The class means $cm$ interfere with the analysis of G-VAE. For example, if the FLN of two music bars are different and other music symbols are the same, the addition of different class means can cause a big difference between the two latent vectors. So in Figure \ref{fig:side:d}, we subtract $cm$ before G-VAE encoding and add $cm$ after L-CVAE sampling.

Finally, There are two ways to specify the FLN sequence of a song which will be added to the input for L-CVAE decoder during the generation phase of HCVAE. One is to specify directly artificially, such as using the FLN sequence of the sample in the dataset. The other is to construct an ordinary VAE to generate it.

\subsection{Hirerarchical Conditional GAN} 
As shown in Figure \ref{fig:side:b}, On the basis of using the WGAN-GP model, we let HCVAE decoder be the generator of WGAN-GP. At the same time, we also share the encoder layer of L-CVAE and put it into the Critor of WGAN-GP in order to better smooth the training curve and solve the problem of mode collapse. In this way, HCGAN obtains the ability of HCVAE conditional constraints and stratification. The training of HCVAE (Task1 and Task2) as a pre-training of HCGAN training (Task3) makes WGAN-GP skip the difficult problem of convergence, and also expands the HCVAE generation effect through HCGAN model. 
\section{Experiment}

\subsection{Experiment Details} 
\subsubsection{Dataset}
The dataset of the paper uses the Nottingham Dataset, which is equivalent to mnist in the field of music generation. The Nottingham Dataset provides music in both MIDI format and ABC format. In our experiment, we selecte the widely used MIDI format music file. The Nottingham Dataset also provides files for MIDI music with a chord version and a clean version without a chord. Since the purpose of this paper is to generate single track music, we select the latter.
\subsubsection{Data Preprocessing}
First, we convert the music to 12 majors, the speed is adjusted to 120bpm. Then, we convert the MIDI file into a piano roll format, which is a 2-dimensional matrix. The horizontal dimension of the matrix represents time, and the vertical dimension represents 128 pitches. For L-CVAE, we split all the songs in the data set into music phrases with a shape of (50, 128) and a length of 16 beats. At the same time, we choose the FLN of each phrase to form a note pair, representing the FLN condition. We add a note pair with frequnecy more than 20 times to the note pair class dictionary, and finally, the dictionary length is 125. These note pairs are labed from 1 to 125, and we add additional two labels: 126th label is as the other unconstrained tonic tag, and 0th label as the empty class tag. So we get total of 127 categories. We convert it to a One-Hot form, as a constraint tag for each phrase, participating in the training of the model.
For HCVAE, We select 17 phrases with the highest frequency of occurrence as the uniform length. Each song only selects the first 17 phrases, and the rest is discarded. The shape of the cut is (850, 128).
\subsubsection{Parameter Details}
The batch size for both HCVAE (Task2) and L-CVAE (Task3) is 32. In L-CVAE, a latent vector size is 32, and the sampling error variance is 0.01. In G-VAE, a latent vector size is 256, and the variance of the sampling error is set to 0.1. For both LSTM of L-CVAE and G-VAE, an intermediate-dim is 256.
The experiment is performed on the NVIDIA P100 with 16G memory, and Python3 and Keras deep learning framework based on the Tensorflow backend are built under the docker image. Complete training can be completed in one day.

\subsection{Result} 
In view of the music generation experiment is difficult to evaluate, we investigate the experimental part of the classic well-known papers in the field of music generation. MuseGAN\cite{dong2018musegan} proposes two evaluation methods: Objective Evaluation and User Study. MidiNet\cite{yang2017midinet} provides a result of the user study and online audio examples.
Xiaoice band\cite{zhu2018xiaoice} uses Human Evaluation to evaluate rhythm, melody, integrity, singability, and uses Chord Accuracy to evaluate chords. Pi from song\cite{chu2016song} provides reports of 27 volunteers voted and provides online samples. Transformer\cite{huang2018music} shows the Validation NLL and Human Evaluation of their models. 

Considering the characteristics of our model, we propose a series of methods to verify the comprehensive quality of the model. We employ to MuseGAN's Objective Evaluation to evaluate the quality of the music generated by the HCVAE and HCGAN. To verify the effectiveness of the First and Last Notes (FLN), we propose FLN Accuracy inspired by Xiaoice's Chord Accuracy. After that, we verify the model from the perspective of length and musical form structure. We also use volunteers to vote on the six indicators of music to verify the quality of the generated music. Finally, for more audio examples generated by MIDI-Sandwich, please go to \url{https://github.com/LiangHsia/MIDI-Sandwich}.

\subsubsection{Objective Evaluation} 
We employ the MuseGAN evaluation method Objective Metrics for evaluation. Since the MIDI-Sandwich generates single-track music, we remove the drum pattern and the tonal distance which are from the original MuseGAN evaluation method. The irregular tone ratio is added as our evaluation indicator. We evaluate the real experimental set (dataset), generation of L-CVAE (Task1), HCVAE (Task2) and MIDI-Sandwich (Task3). The result is shown in Table \ref{tab:aStrangeTable1}.

\begin{table}[!htbp]
	\vspace*{-0.5\baselineskip}
	\centering
	\caption{Objective Metrics Evaluation(EB: ratio of empty bars(in \%), UPC: number of used pitch classes per bar (from 0 to 12), QN: ratio of "qualified" notes, IT: ratio of irregular tone (in \%)).}\label{tab:aStrangeTable1}
	\begin{tabular}{c|c|c|c|c}
		\hline
		& empty bars & used pitch classes& qualified notes& irregular tone\\
		& (EB: \%) & (UPC) & (QN: \%) & (IT: \%) \\
		\hline\hline
		Dataset & \textbf{8.06} & \textbf{1.71} & \textbf{90.0} & \textbf{0.00}\\
		\hline\hline
		from L-CVAE  & 6.22 & 3.64 & 63.0 & 6.52\\
		from HCVAE  & 5.92 & 3.28& 71.2 & 6.13\\
		from MIDI-Sandwich & \textbf{6.71} & \textbf{2.67}& \textbf{76.7} & \textbf{2.12}\\
		\hline
	\end{tabular}
	
\end{table}

As can be seen from the Table \ref{tab:aStrangeTable1}, music generated by MIDI-Sandwich is more fluent than that by HCVAE and L-CVAE, and the probability of crowded notes and irregular sounds is reduced. It  proves that MIDI-Sandwich which combines HCVAE and HCGAN has better effect than a single model.

\subsubsection{Length Comparison} 

\begin{table}[!htbp]
	
	\centering
	\caption{Contrast Length}\label{tab:aStrangeTable2}
	\begin{tabular}{cccccccc}
		\hline
		& Pi& Xiaoice & MusicVAE &MIDI-GAN &Musegan &MagentaRNN &MIDI-Sandwich\\
		\hline
		Length/bar& 8& 8& 16& 8& 8& 8& \textbf{34}\\
		\hline
	\end{tabular}
	\vspace*{-0.5\baselineskip}
\end{table}

As can be seen from the Table \ref{tab:aStrangeTable2}, MIDI-Sandwich is compared with the classic model and the popular models in recent years. Our model length is 34 bars (34x4 beats). It can be seen that our model generates the longest length of valuable music sequences. Compared with MusicVAE and Transormer, MIDI-Sandwich has a conditional constraint mechanism and introduces the knowledge of music theory such as musical form, melodic motion and tonic, which makes the melody more directional and closer to human standards of appreciation.

\subsubsection{Musical Form}
We prove the structural validity of the music sequence generated by this model from the perspective of comparing the latent vectors.

\begin{figure}
	\begin{minipage}[t]{0.48\textwidth}
		\centering
		\includegraphics[width=1.4in]{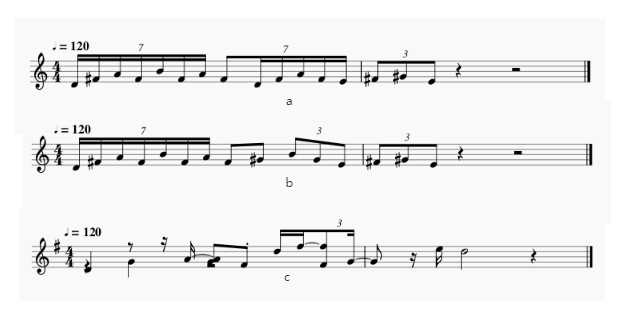}
		\caption{Three music fragments within a song that have relationships.}
		\label{fig:side:a1}
	\end{minipage}%
	\begin{minipage}[t]{0.48\textwidth}
		\centering
		\includegraphics[width=1.4in]{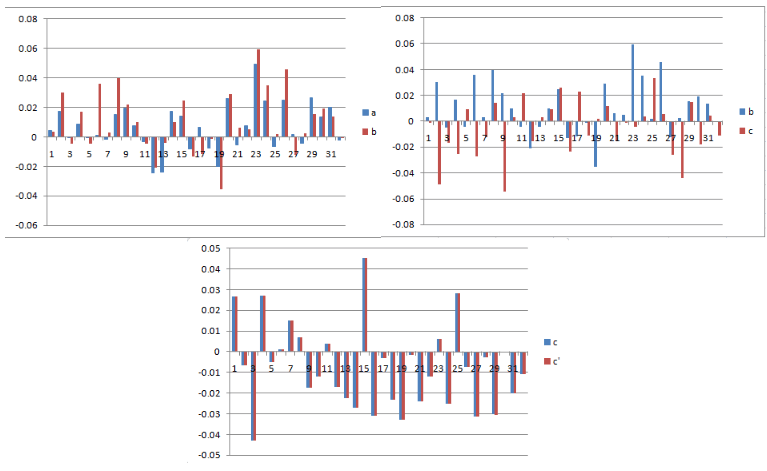} 
		\caption{The comparison of different structural relationships of music at latent.
			Top left: ab similar, top right: bc sandhi, bottom: cc' repeat.}
		\label{fig:side:b1}
	\end{minipage}
\vspace*{-1.5\baselineskip}
\end{figure}

As shown in Figure \ref{fig:side:a1}, we extract three fragments in the same song: a, b, and c. Figure \ref{fig:side:b1} prestens L-CVAE latent vector of three fragments (after subtracting the $cm$), we can observe that the structural relationship at the music level can be expressed at the latent vector level. That is, the difference in the dimensions of latent vector is less than one thousandth for same music fragments. The dimensions of latent vector is mostly similar for similar music fragments. And the difference of latent vector dimensions between the two different music fragments is also significant.

\begin{figure}
	\begin{minipage}[t]{0.48\textwidth}
		\centering
		\includegraphics[width=1.4in]{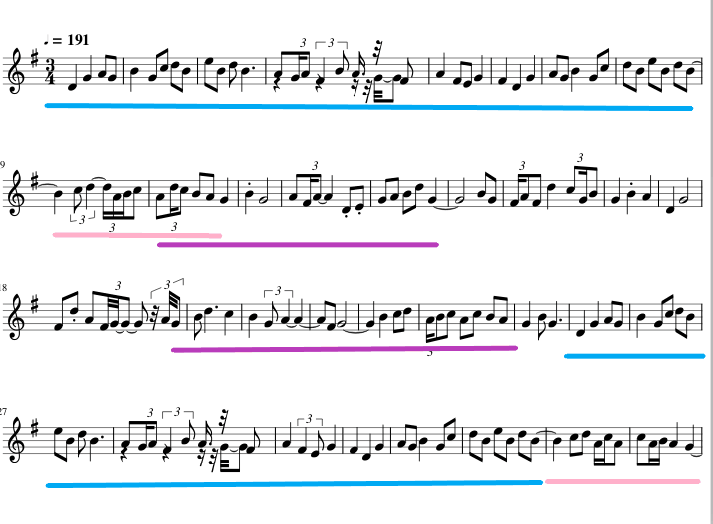}
		\caption{Long-term music clips generated by MIDI-Sandwich.}
		\label{fig:side:b11}
	\end{minipage}%
	\begin{minipage}[t]{0.48\textwidth}
		\centering
		\includegraphics[width=1.4in]{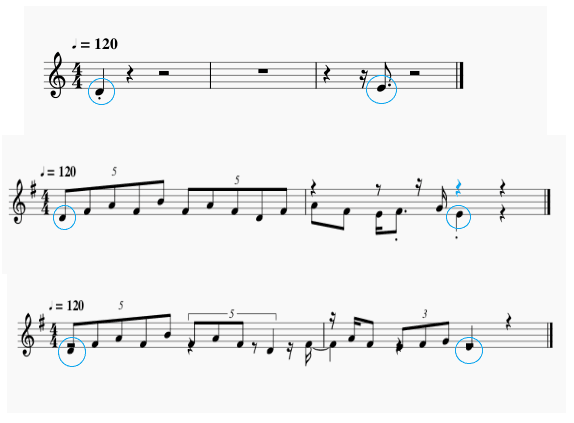} 
		\caption{Three music fragments.}
		\label{fig:side:b22}
	\end{minipage}
\vspace*{-1.5\baselineskip}
\end{figure}

As shown in the Figure \ref{fig:side:b11}, it is a music score generated by the MIDI-Sandwich. the music clip on the blue line reflects the repeating structure of the paragraph.
The clip on the pink line has a slight change after the same repeating music segment. The fragments on the purple line reflect the rise and fall of the melody. They present that the music generated by our model is composed of repeating, similar, transposing musical form structures.
\vspace*{-1.2\baselineskip}

\subsubsection{The First and Last Notes} 
There are three music fragments in Figure \ref{fig:side:b22}: a sample of the first and last notes (FLN) and two music fragments generated based on this sample constrains.
It can be clearly seen that the generated music fragments satisfy the constraint and also have diversity. 
FLN Accuracy method is used to evaluate the accuracy of generating music bars to obey the FLN constraint. The result shows the accuracy can reach 94\%, proving that the majority of generated music fragments almost satisfy the constraint of the FLN, which leads generated musics have stable tonal performance and melodic motion.

\subsubsection{User Study}

\begin{table}[!htbp]
\vspace*{-2\baselineskip}
	\centering
	\caption{User Study (R: Rhythm, M: Melody, I: Integrity, Si: Singability, St: Structrue, MM: Melodic Motion.)}\label{tab:aStrangeTable3}
	\begin{tabular}{c|cccccc|c}
	\hline\hline
	Methods   & R& M& I& Si &St & MM &Average\\
	\hline
	Magenta-RNN& \textbf{3.42}& \textbf{3.52} &3.23 &\textbf{2.57} &2.90 &2.95 &3.10\\
	MidiNet(GAN)& 3.09& 3.19 &2.38 &2.14 &2.47 &3.09 &2.56\\
	VAE-GAN &1.90& 2.33 &2.52 &2.09 &2.23 &2.42 &2.25\\
	\hline
	Sandwich(HCVAE)& 2.52& 3.14 &3.28 &2.14 &3.57 & 3.42&3.01\\
	Sandwich(HCVAE-GAN)& 2.76& 3.38 &\textbf{3.33} &2.28 &\textbf{3.66} & \textbf{3.47}&\textbf{3.15}\\
	\hline\hline
	\end{tabular}
\vspace*{-0.5\baselineskip}	
\end{table}

We randomly invited 21 volunteers online to compare songs generated by MIDI-Sandwich, Magenta-RNN, MIDI-Net, and VAE-GAN. As shown in the Table \ref{tab:aStrangeTable3}, we asked them to rate these songs from 1 to 5 for 6 indicators. The evaluation shows that MIDI-Sandwich performs well on both structrue, melodic motion indicators and average of songs. Melody, integrity, singability are similar to others generated by other models. Our model is slightly weak in rhythm. 

\section{Conclusion}
In this paper, we propose a multi-task multi-model hierarchical conditional VAE-GAN network, named MIDI-Sandwich. The MIDI-Sandwich has four models (L-CVAE, G-VAE, HCVAE, HCGAN) and three tasks to product a long-term symbol single-track music. At first, we train the L-CVAE (Task 1), to generate music fragments which are pre-specified the First and Last Notes (FLN) constraint. Then, We train HCVAE (Task2), which consists of L-CVAE and G-VAE, to generate songs. The underlying L-CVAE deals with the local music fragments. And the upper G-VAE controls global structure and the melody direction of the song by analyzing latent vector sequences from L-CVAE encoder and providing L-CVAE decoder with the FLN sequences. After HCVAE training, the decoder of HCVAE is regarded as the generator of HCGAN, and some parts of HCVAE encoder and CNN layer are used as the critor of HCGAN. Then we train HCGAN (Task3) to further improve the quality of generating songs. 

Through experiments, they prove that songs generated by MIDI-Sandwich have three advantages. First, the MIDI-Sandwich directly outputs a song which has a long length instead of a music fragment. Second, The MIDI-Sandwich makes the generated music have a musical form structure by layered network architecture. Finally, the FLN is used instead of the chord progression which is commonly used in multi-track music as a constraint, which has low requirements on the data format and enhances the applicability of the networks. Moreover, the FLN constraints also keep the melody stable in tonality. At the same time, the MIDI-Sandwich uses a multi-model and multi-task architecture. Each sub-model can be flexibly expanded or replaced. Each sub-task can also be easily evaluated. In the future, we will continue to improve the network model in both rhythm and multi-track, and try to use the model in other artificial intelligence symbol music fields.

\bibliography{ref}
\bibliographystyle{splncs04}
%
%
%
%
%
\end{document}